\documentclass[preprint]{aastex63}
\usepackage{comment}
\usepackage{bm}
\usepackage{float}
\usepackage{multirow}
\usepackage{txfonts}
\graphicspath{{./}}
\received{June 26, 2021}
\revised{February 22, 2022}
\accepted{March 7, 2022}
\submitjournal{Astronomical Journal}

\shorttitle{Size Distribution of L${}_5$ Jupiter Trojans}
\shortauthors{Uehata, Terai, Ohtsuki et al.}

\begin{document}

\title{Size Distribution of Small Jupiter Trojans in the L${}_5$ Swarm
\footnote{
This research is based on data collected at Subaru Telescope, which is operated by the National Astronomical Observatory of Japan.
We are honored and grateful for the opportunity of observing the Universe from Maunakea, which has the cultural, historical and natural significance in Hawaii.
} 
}

\correspondingauthor{Keiji Ohtsuki}
\email{ohtsuki@tiger.kobe-u.ac.jp}

\author{Kotomi Uehata}
\affiliation{Department of Planetology, Kobe University, Kobe 657-8501, Japan}
\affiliation{NEC Corporation, Tokyo 108-8001, Japan}

\author[0000-0003-4143-4246]{Tsuyoshi Terai}
\affiliation{Subaru Telescope, National Astronomical Observatory of Japan,
National Institutes of Natural Sciences, Hilo, HI 96720, USA}

\author[0000-0002-4383-8247]{Keiji Ohtsuki}
\affiliation{Department of Planetology, Kobe University,
Kobe 657-8501, Japan}

\author[0000-0002-3286-911X]{Fumi Yoshida}
\affiliation{Planetary Exploration Research Center, Chiba Institute of Technology, Narashino 275-0016, Japan}
\affiliation{School of Medicine, University of Occupational and Environmental Health, 
Kita-Kyushu, Japan}








\begin{abstract}

We present an analysis of survey observations of the trailing L${}_5$ Jupiter Trojan swarm using the wide-field Hyper Suprime-Cam CCD camera on the 8.2 m Subaru Telescope.
We detected 189 L${}_5$ Trojans from our survey that covered about 15 deg${}^2$ of sky with a detection limit of $m_r = 24.1$ mag, and selected an unbiased sample consisting of 87 objects with absolute magnitude $14 \lesssim H_r \leq 17$ corresponding to diameter $2 \ {\rm km} \lesssim D \lesssim 10 \ {\rm km}$ for analysis of size distribution.
We fit their differential magnitude distribution to a single-slope power-law with an index  $\alpha = 0.37 \pm 0.01$, which corresponds to a cumulative size distribution with an index of $b = 1.85 \pm 0.05$.
Combining our results with data for known asteroids, we obtained the size distribution of L${}_5$ Jupiter Trojans over the entire size range for $9 \lesssim H_V \leq 17$, and found that the size distributions of the L${}_4$ and L${}_5$ swarms agree well with each other for a wide range of sizes.
This is consistent with the scenario that asteroids in the two swarms originated from the same primordial population. 
Based on the above results, the ratio of the total number of asteroids with $D \geq 2$ km in the two swarms is estimated to be $N_{\rm L4}/N_{\rm L5}=1.40 \pm 0.15$, and the total number of L${}_5$ Jupiter Trojans with $D \geq 1 {\rm km}$ is estimated to be $1.1 \times 10^5$ by extrapolating the obtained distribution.

\end{abstract}




\section{INTRODUCTION} \label{sec:intro}

The Jupiter Trojans are a group of asteroids that share their orbits with Jupiter and are confined to two swarms centered about the L${}_4$ and L${}_5$ Lagrangian points, which lead and trail Jupiter by about 60 degrees.
Although the gravitational effects of the giant planets likely have reduced the number of the swarm asteroids over time, most of Jupiter Trojans have stability timescales comparable to or longer than the age of the Solar System \citep{le97,di14}.
While in situ formation of Jupiter and the Trojans at 5.2 au cannot explain their characteristics such as broad inclination distribution, recent models showed that capture of icy planetesimals from the trans-Neptunian region at the time of orbital instability of giant planets can explain their observed orbital characteristics and estimated total mass \citep{mo05,ne13}. 
On the other hand, a more recent model of capture of Trojans asteroids due to the growth and inward migration of Jupiter shows a higher capture efficiency than the above model based on orbital instability, and can naturally explain the observation that the L${}_4$ swarm is more populated than the L${}_5$ swarm \citep{pi19}.
In the latter model, Trojans are originated from the radial location where Jupiter's core formed.
Thus, Trojan asteroids are expected to provide us with clues to understanding origin and radial transport of small icy bodies caused by orbital evolution of giant planets as well as building blocks of the planets \citep[see a review by][]{em15}.

Trojan asteroids are expected to hold unique information about primitive material in the outer Solar Sytem region, and detailed information for some of these asteroids is expected to be revealed by NASA's Lucy mission \citep{le21}.
On the other hand, in order to collect information from a large number of samples for statistical investigation, various observations using ground-based or space telescopes have been carried out \citep{em15}.
Here we focus on observation of size distribution of Jupiter Trojans, which can provide us with clues to their origin and collisional evolution.
Size distribution of Trojan asteroids has been investigated by several surveys \citep[e.g.][]{je00, sz07}.
The slopes of the size distributions of hot Kuiper-belt objects (KBOs) and Jupiter Trojans near their large size end (with absolute magnitudes $H \lesssim 8$, corresponding to diameters $D \gtrsim 100$ km) are shown to be similar but is distinctly shallower than that of cold KBOs \citep{fr14}.
This can be interpreted as a support for the view that Jupiter Trojans were captured from scattered KBOs \citep{fr14}, but information for smaller objects will be useful to derive stronger constraints.
Size distribution of small Trojan asteroids down to $\sim$ 2 km has also been examined using the Suprime-Cam CCD camera on Subaru Telescope for the L${}_4$ \citep{yn05,wo15} and L${}_5$ swarms \citep{yn08}, respectively.
It is also known that Jupiter Trojans have two groups with respect to their color or albedo \citep[e.g.][]{sz07,em11,gr12}, and each group of asteroids are shown to have distinct size distribution \citep{wo14,wo15}.
However, in the case of small Trojans in the L${}_5$ swarm, the number of objects used for the analysis of size distribution
was limited (62 in \citet{yn08}) and, furthermore, removal of observational biases may have not been sufficient (\S 4.1).

On the other hand, a new CCD camera for Subaru Telescope with a larger field of view, called Hyper Suprime-Cam (HSC), has been developed in 2012 \citep{mi12} and became a general user instrument in 2014.
Using HSC, we have carried out observational studies of size distributions of L${}_4$ Jupiter Trojans \citep{yt17} and Hilda asteroids \citep{ty18}, colors of KBOs \citep{te18} and Centaurs \citep{sa18}, color and size distribution of main-belt asteroids \citep{ma21}, and comparison of size distributions of small bodies in various populations (\cite{yo19}, see also \cite{yo20} for corrections) .

As an extension of this series of works, in the present work, we examine size distribution of small L${}_5$ Jupiter Trojans with HSC. By comparing our results with \citet{yt17}, who examined size distribution of small L${}_4$ Jupiter Trojans also using HSC, we will discuss implications for the origin of Jupiter Trojans.
The organization of this paper is as follows.
\S 2 describes our observation and data analysis, and results are presented in \S 3.
Discussion is given in \S 4, where we compare our results with previous studies on size distribution of Jupiter Trojans in detail.
Our conclusion is given in \S 5.


\section{Observation and Data Analysis}

\begin{figure}
\begin{center}
 \includegraphics[width=12cm,pagebox=cropbox,clip]{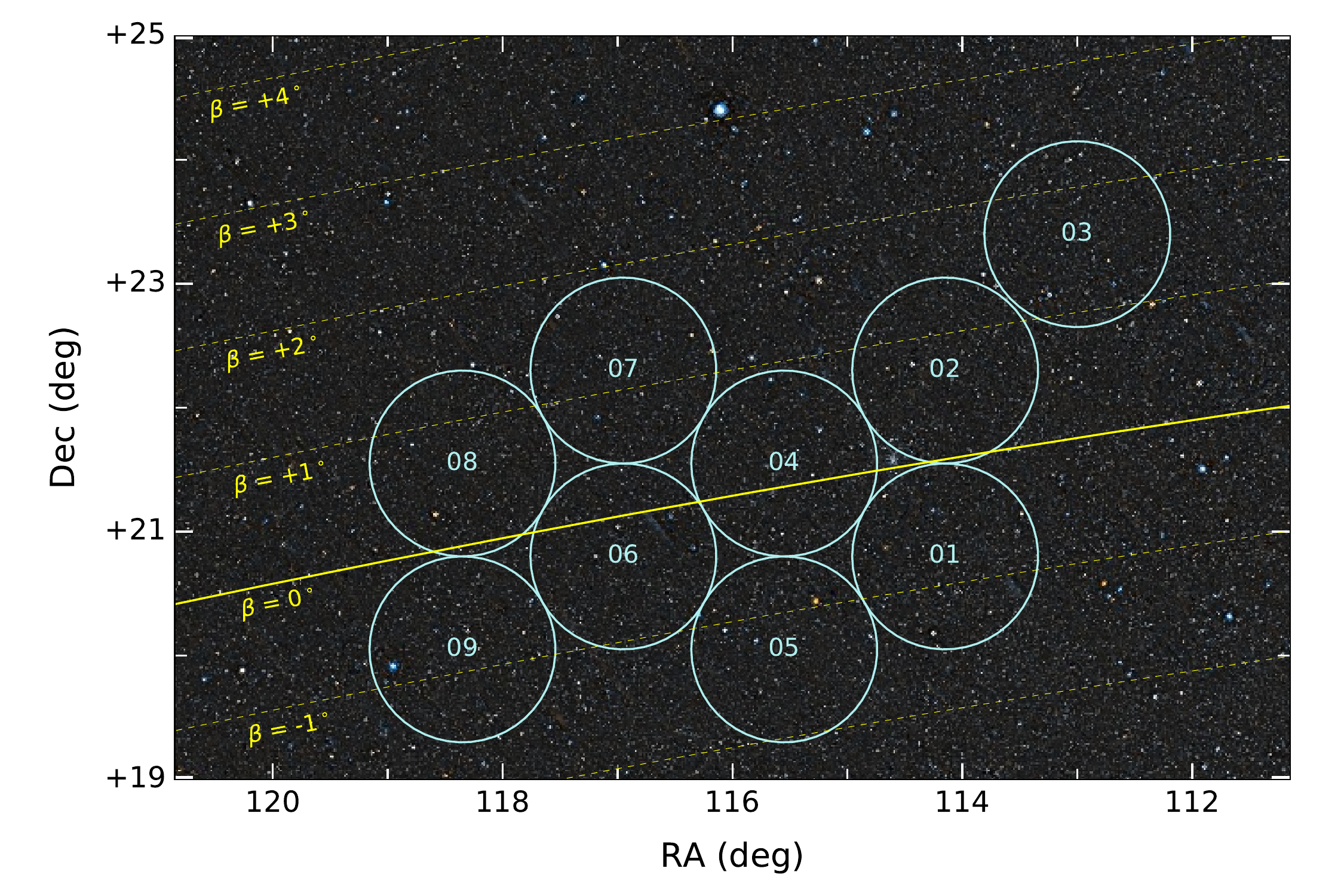}
 \end{center}
 \caption{
 Locations of the nine observed fields of our survey.
 The size of each circle corresponds to the field of view of Hyper Suprime-Cam.
 Lines of ecliptic latitude ($\beta$) are also shown, with the ecliptic plane shown with the thick line.
The background image is taken from Pan-STARRS1 \citep{ch16}.
} 
\label{fig:f1}
\end{figure}

Observation of the L${}_5$ Trojan swarm of Jupiter was carried out on January 10, 2016 (UT) using the HSC installed at the prime focus of the 8.2 m Subaru Telescope atop Maunakea, Hawaii. 
The HSC is a gigantic mosaic camera consisting of 104 CCD chips that can be used for scientific observation with a field of view of 1.5 deg in diameter.
We surveyed about 15 deg${}^2$ of sky in the L${}_5$ region near opposition (Figure \ref{fig:f1}).
The right ascensions (R.A.) of the L${}_5$ Lagrangian point and opposition at our observation were 103 deg and 111 deg, respectively, and the surveyed area was near the L${}_5$ point and was within about 8 deg from opposition.
Nine fields of view of HSC shown in Figure \ref{fig:f1} were observed in the $r$-band, and
the exposure time for each field was 240 seconds. 
Each field of view was visited three times with a time interval of $\sim 23$ minutes.
Observational sequence is shown in Table \ref{table:t1}.
The average seeing size of each field was between $0 \farcs 82$ and $1 \farcs 19$, allowing us observations under a rather stable condition.

\begin{table}[h]
\centering
\caption{Observation Log}

\begin{tabular}{ccccccc} \hline
Field \ ID & UTC & Time interval & R.A. & Decl. & Airmass & Seeing \\
            &        & (min: sec) & (deg) & (deg) &            & (arcsec) \\ \hline
FIELD01 & 06:35:33 & $-$ & 114.14959 & 20.80001 & 1.859 & 1.23 \\
            & 06:58:19 & 22:46 & 114.14958 & 20.80000 & 1.633 & 1.13 \\
            & 07:21:01 & 22:42 & 114.14960 & 20.80000 & 1.466 & 1.20 \\
FIELD02 & 06:40:06 & $-$ & 114.14959 & 22.30001 & 1.796 & 1.19 \\
            & 07:02:51 & 22:45 & 114.14959 & 22.30003 & 1.588 & 1.05 \\
            & 07:25:35 & 22:44 & 114.14959 & 22.30001 & 1.433 & 1.23 \\
FIELD03 & 06:44:38 & $-$ & 112.99960 & 23.40003 & 1.698 & 1.27 \\
            & 07:07:23 & 22:45 & 112.99959 & 23.40001 & 1.516 & 1.04 \\
            & 07:30:08 & 22:45 & 112.99961 & 23.40003 & 1.379 & 1.19 \\
FIELD04 & 06:53:45 & $-$ & 115.54960 & 21.54999 & 1.720 & 1.20 \\
            & 07:16:28 & 22:43 & 115.54960 & 21.54999 & 1.531 & 1.15 \\
            & 07:39:14 & 22:46 & 115.54958 & 21.55000 & 1.389 & 1.14 \\
FILED05 & 08:12:45 & $-$ & 115.54958 &  20.05001 & 1.241 & 0.90 \\
            & 08:35:35 & 22:50 & 115.54960 & 20.05000 & 1.167 & 0.81 \\
            & 08:58:24 & 22:49 & 115.54960 & 20.05000 & 1.109 & 0.80 \\
FIELD06 & 08:17:19 & $-$ & 116.94958 & 20.80002 & 1.244 & 0.84 \\
            & 08:40:09 & 22:50 & 116.94958 & 20.80002 & 1.169 & 0.83 \\
            & 09:03:03 & 22:54 & 116.94960 & 20.80002 & 1.111 & 0.79 \\
FIELD07 & 08:21:53 & $-$ & 116.94958 & 22.30004 & 1.226 & 0.86 \\
            & 08:44:42 & 22:49 & 116.94960 & 22.30002 & 1.155 & 0.87 \\
            & 09:07:35 & 22:43 & 116.94960 & 22.30004 & 1.101 & 0.75 \\
FIELD08 & 08:26:26 & $-$ & 118.34958 & 21.55001 & 1.230 & 0.90 \\
            & 08:49:18 & 22:52 & 118.34960 & 21.54999 & 1.158 & 0.91 \\
            & 09:12:14 & 22:56 & 118.34959 & 21.55003 & 1.103 & 0.75 \\
FIELD09 & 08:31:00 & $-$ & 118.34960 & 20.04999 & 1.216 & 0.91 \\
            & 08:53:51 & 22:51 & 118.34960 & 20.05001 & 1.147 & 1.00 \\
            & 09:16:47 & 22:56 & 118.34958 & 20.05000 & 1.094 & 0.88 \\ \hline
\end{tabular}

\label{table:t1}
\end{table}

\begin{figure}
\begin{center}
 \includegraphics[width=10cm,pagebox=cropbox,clip]{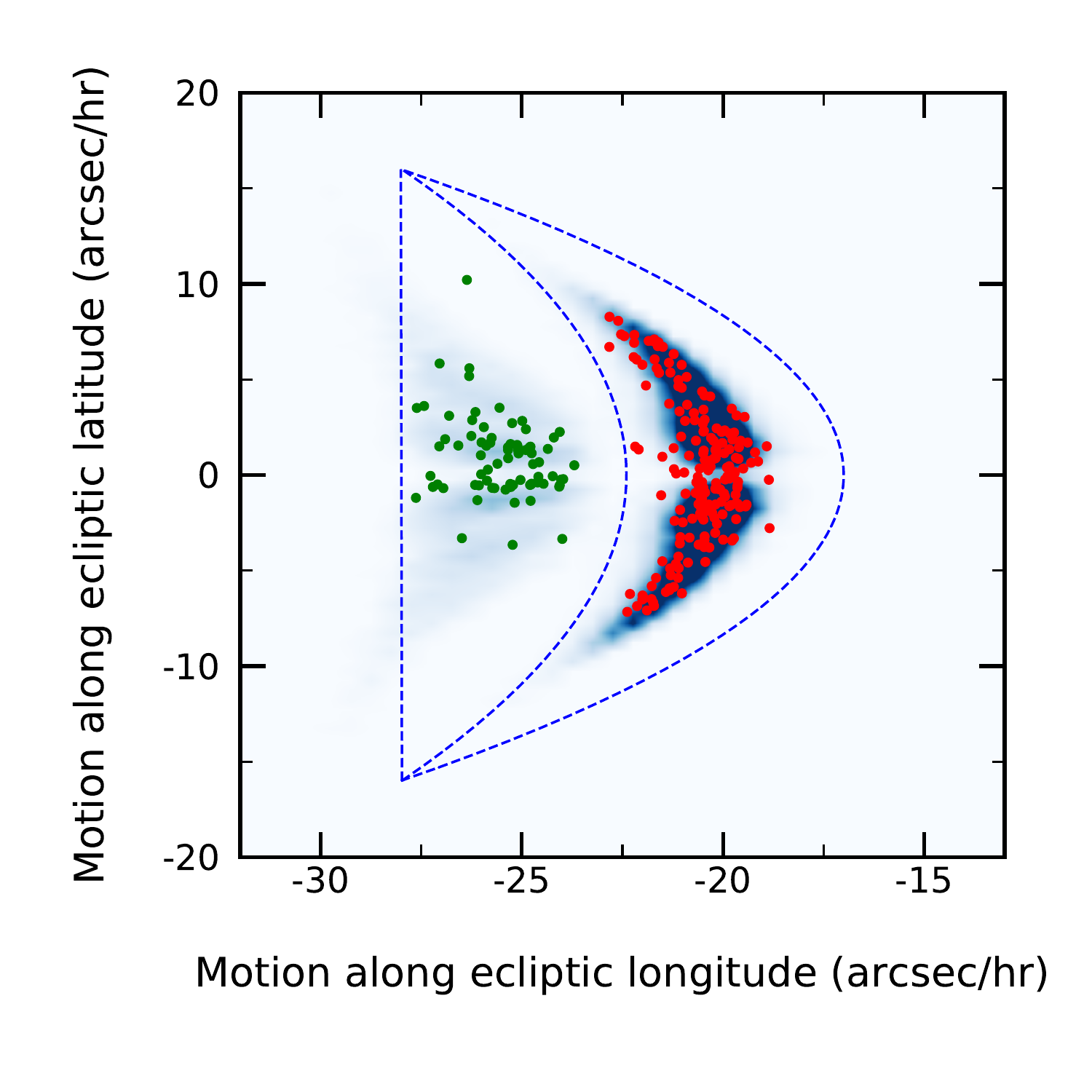}
 \end{center}
 \caption{
 Distribution of apparent velocities of detected Jupiter Trojans (red dots) and Hildas (green dots) along ecliptic longitude and latitude.
The background map shows the distribution of apparent velocities of synthetic Trojans and Hildas, respectively, generated by a Monte Carlo method based on the orbital distribution of known asteroids in each population.
The three dashed lines show boundaries used to distinguish detected objects for each population.
} 
\label{fig:f2}
\end{figure}

Data analysis was performed following procedures similar to those described in \citet{yt17} and \citet{ty18}.
First, the data are processed with the HSC data reduction$/$analysis pipeline \verb+hscPipe+ (version 3.8.5; \citet{bo18}).
We extracted moving object candidates from the source catalogs created by \verb+hscPipe+.
Then we identify L${}_5$ Jupiter Trojans based on their apparent motion \citep[see][for details of the procedures]{yt17}.
Figure \ref{fig:f2} shows the velocities along the ecliptic longitude and latitude of the detected Jupiter Trojan (red) and Hilda (green) candidates.
We searched candidate objects for the two populations within the outer boundary shown by the left vertical line and the right curve, and the middle curve was used to distinguish between Hildas and Trojans.
We can see that the two populations are clearly separated from each other.
From these procedures, we identified 189 L${}_5$ Jupiter Trojans and 76 Hildas from our sample.
We analyzed the data for the Jupiter Trojans in the present work.

For each of the detected Trojans, we carried out photometric measurements as follows (see \citet{yt17} and \citet{ty18} for details).
First, we determined the center position for each of the detected objects using $\chi^2$-fitting of an object model to the image data; the model was generated by integration of Gaussian profiles with the center shifted with a constant motion corresponding to the measured velocity.
Then we measure the total flux for each object by aperture photometry \citep{yt17,ty18}.
The measured flux was converted into apparent magnitude in the AB magnitude system using the photometric zero point, which was estimated in \verb+hscPipe+ based on the data of the Panoramic Survey Telescope and Rapid Response System (Pan-STARRS) 1 survey \citep{sc12,to12,ma13}.
Photometric measurements were performed for 187 of the detected objects, 
excluding two objects for which the measurement failed due to either a background bright star or a defect in part of the images.

\begin{figure}
\begin{center}
\includegraphics[width=15cm,pagebox=cropbox,clip]{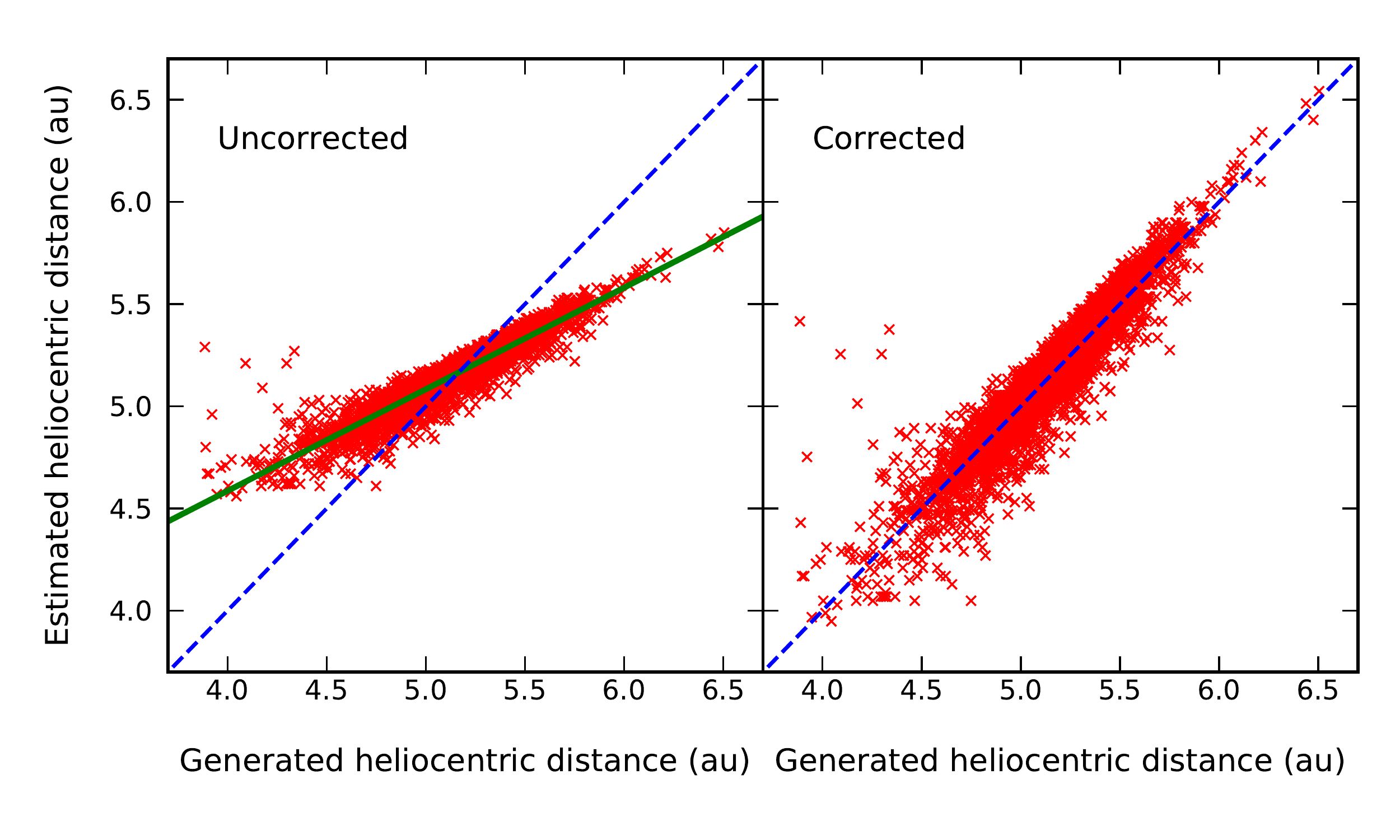}

\caption{
Left: Relationship between true generated heliocentric distance of synthetic Trojans and the values of the heliocentric distance estimated from their apparent motion under the same observation conditions as our survey.
The systematic deviation between the two values is due to the assumption of circular orbits in the estimation from apparent motions.
The solid line represents the best-fit to the estimated values given by Eq.(\ref{eq:helio_cor}).
Right: Relationship between the generated heliocentric distance and re-estimated values.
The latter was obtained by removing the systematic deviation using Eq.(\ref{eq:helio_cor}).
}
\label{fig:f3}
\end{center}
\end{figure}

The short observation arcs of the detected objects did not allow us to determine their orbits accurately.
However, because the surveyed area was close to opposition, assuming that their orbits are circular we were able to estimate the heliocentric distance and orbital inclination of each of the detected objects from its apparent motion.
We estimated the heliocentric distance and orbital inclination following \citet[][see also \citet{je96}]{te13}.
In this case, systematic deviation from the correct value is expected in the estimated heliocentric distance because of the assumption of circular orbits.
This systematic errors in estimated heliocentric distance can be corrected as follows \citep{yt17,ty18}.
First, we generate synthetic orbits of Jupiter Trojans from probability distributions based on the orbital distributions of known Trojans obtained from Minor Planet Center (MPC) database.
\footnote{http://minorplanetcenter.net/iau/lists/JupiterTrojans.html}
Then their apparent motions with the same conditions as our observations are calculated, and their heliocentric distances are estimated from their apparent motions in the manner described above.
The left panel of Figure \ref{fig:f3} shows a comparison between their estimated heliocentric distance ($R_{\rm est}$)  and the true generated values ($R_{\rm true}$), where the systematic deviation between the two values can be seen.
The solid line represents the best-fit linear function to the obtained distribution given as

\begin{eqnarray}
R_{\rm est} = 0.497 R_{\rm true} + 2.596.
\label{eq:helio_cor}
\end{eqnarray}

\begin{figure}
\begin{center}
 \includegraphics[width=11cm]{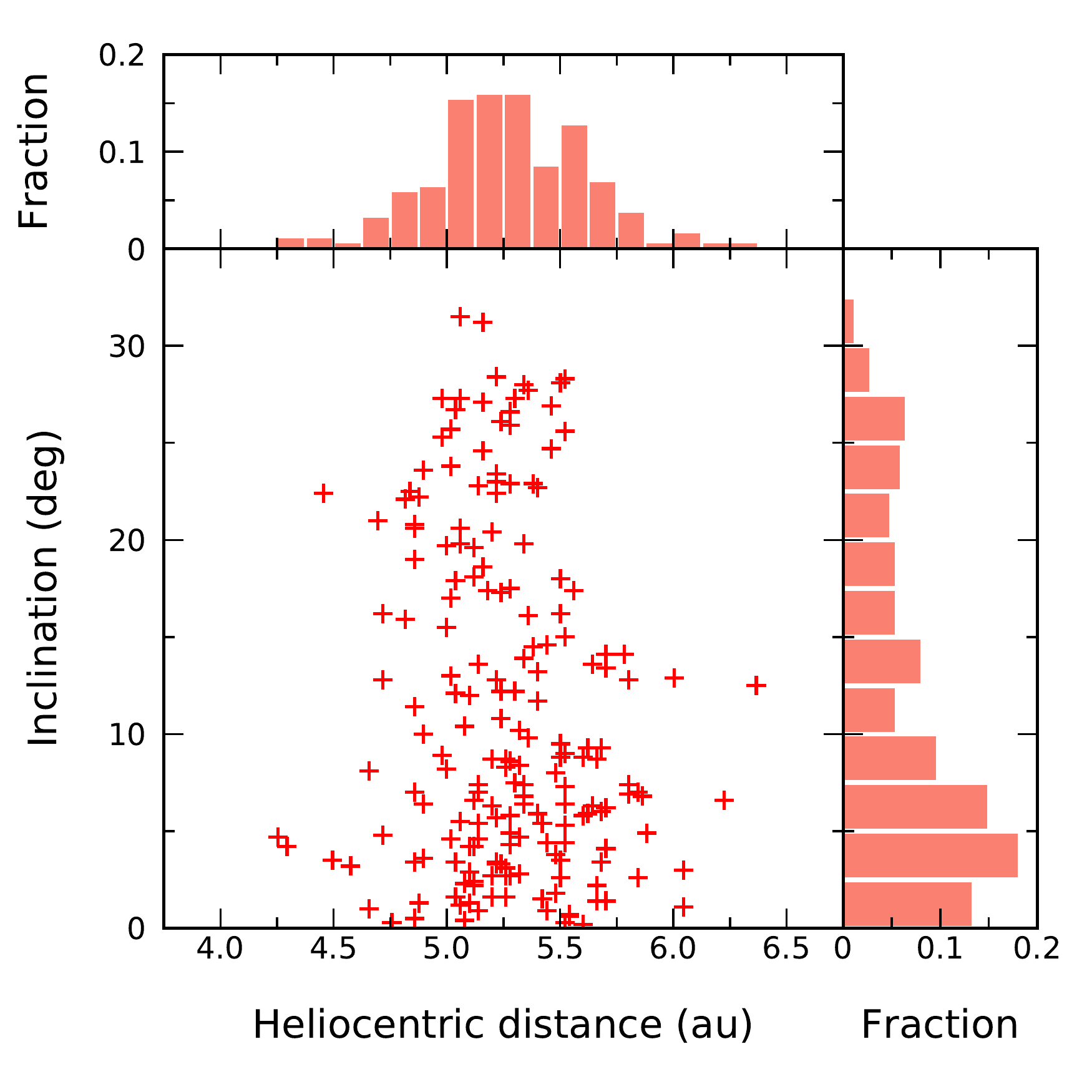} 
 \end{center}
 \caption{
Relationship between the derived heliocentric distances and orbital inclinations together with their histograms for the detected 189 
Trojan asteroids.
Random errors for the obtained heliocentric distances and orbital inclinations are 0.18 au and 1.5 deg, respectively.
} 
\label{fig:f4}
\end{figure}

\noindent
Using this derived relationship, we re-estimated the heliocentric distance of all the synthetic objects and confirmed that their true values can be recovered (Figure \ref{fig:f3}, right panel).
The statistical error of the corrected heliocentric distance was about 0.09 au \citep{yt17}.
Figure \ref{fig:f4} shows the relationship between the derived heliocentric distance and orbital inclination together with their histograms for our 189 Trojan asteroids.
Random errors for the obtained heliocentric distances and orbital inclinations are 0.18 au and 1.5 deg, respectively.

Detectability of small objects such as Trojan asteroids depends on their apparent magnitudes as well as image quality.
Therefore, for an accurate investigation of these bodies, evaluation of detection completeness is important for removal of such biases.
In the case of HSC, image quality as well as effective area for detection of moving objects differs for each CCD, and the detection efficiency also depends on the observation condition for each visit of the field.
In the present work, we examined the detection efficiency by implanting synthetic moving objects into the image obtained for each field and for each visit, and processing them with  \verb+hscPipe+ in the same manner as for the actual detected moving objects.
Detailed procedures are similar to those described in \citet[][their \S 3.3 and Figure 5]{yt17}, but
here we improved the method by implanting synthetic objects into actual processed images rather than synthetic blank images as done in \citet{yt17}.
This improvement allowed us more accurate estimate of the detection efficiency for each image used for our size distribution analysis.
The obtained detection efficiency as a function of apparent magnitude $m$ is then fit by the following function \citep{yt17}:

\begin{eqnarray}
\eta (m) &=& \eta_0 \sum_{k=1,2}^{} \frac{\epsilon_k(m)}{2} \left[ 1- \tanh \left( \frac{m-m_{50}}{w_k} \right) \right] 
\label{eq:eta}
\end{eqnarray}

\noindent
where

\begin{equation}
\begin{array}{ll}
\epsilon_1(m) &= 1- \epsilon_2(m) \\
                  &{= \displaystyle \frac{1}{2} \left[ 1- \tanh \left( \frac{m-m_{50}}{0.2} \right) \right] }.
\end{array}
\label{eq:ep}
\end{equation}

\vspace{2ex}
\noindent
In the above, $\eta_0$ is the maximum detection efficiency, $m_{50}$ is the apparent magnitude where the efficiency is $\eta_0/2$, and $w_k$ $(k = 1, \ 2)$ are the transition widths; all of these are evaluated by fitting of the calculated detection efficiencies of the implanted synthetic objects.
We examined the detection efficiency for CCDs representative of different locations of each field (i.e., near the center, near the edge, and at an intermediate region), and found that the detection efficiency for almost all the CCDs exceeds 0.5 (i.e., more than 50\% detection) when the apparent magnitude of the implanted synthetic object is 24.1 mag.
Thus, we set 24.1 mag as the detection limit of our survey.




\newpage
\section{Results}

\subsection{Sample Selection}

In order to derive size distribution of detected Trojans, the measured apparent magnitude for each object was converted into the absolute magnitude in the $r$-band as \citep{bo89}

\begin{eqnarray}
\label{eq:H}
H_r = m_r - 5 \log (R \Delta) - P(\theta),
\end{eqnarray}

\noindent 
where $m_r$ is the apparent magnitude in the $r$-band; $R$ and $\Delta$ are the heliocentric and geocentric distances, respectively; and $P(\theta)$ is the phase function for a solar phase angle $\theta$ given by

\begin{equation}
P(\theta) = -2.5\log \left[ (1-G) \Phi_1 + G \Phi_2 \right].
\end{equation}

\noindent
Here, $G$ is the slope parameter, and $\Phi_1$ and $\Phi_2$ are the phase functions given by 

\begin{equation}
\Phi_i = \exp (-A_i [ \tan (\theta/2)]^{B_i}) \quad (i =1, \ 2) 
\end{equation}

\noindent
with $A_1 = 3.33$, $A_2 = 1.87$, $B_1 = 0.63$, and $B_2 = 1.22$.
Because the observations were carried out only for one night, we were not able to determine $P(\theta)$ from our observations.
Thus,  following previous works \citep[e.g.][]{gr11,yt17}, we assumed $G=0.15$, which is a typical value for main-belt asteroids.

\begin{figure}
\begin{center}
 \includegraphics[height=10cm]{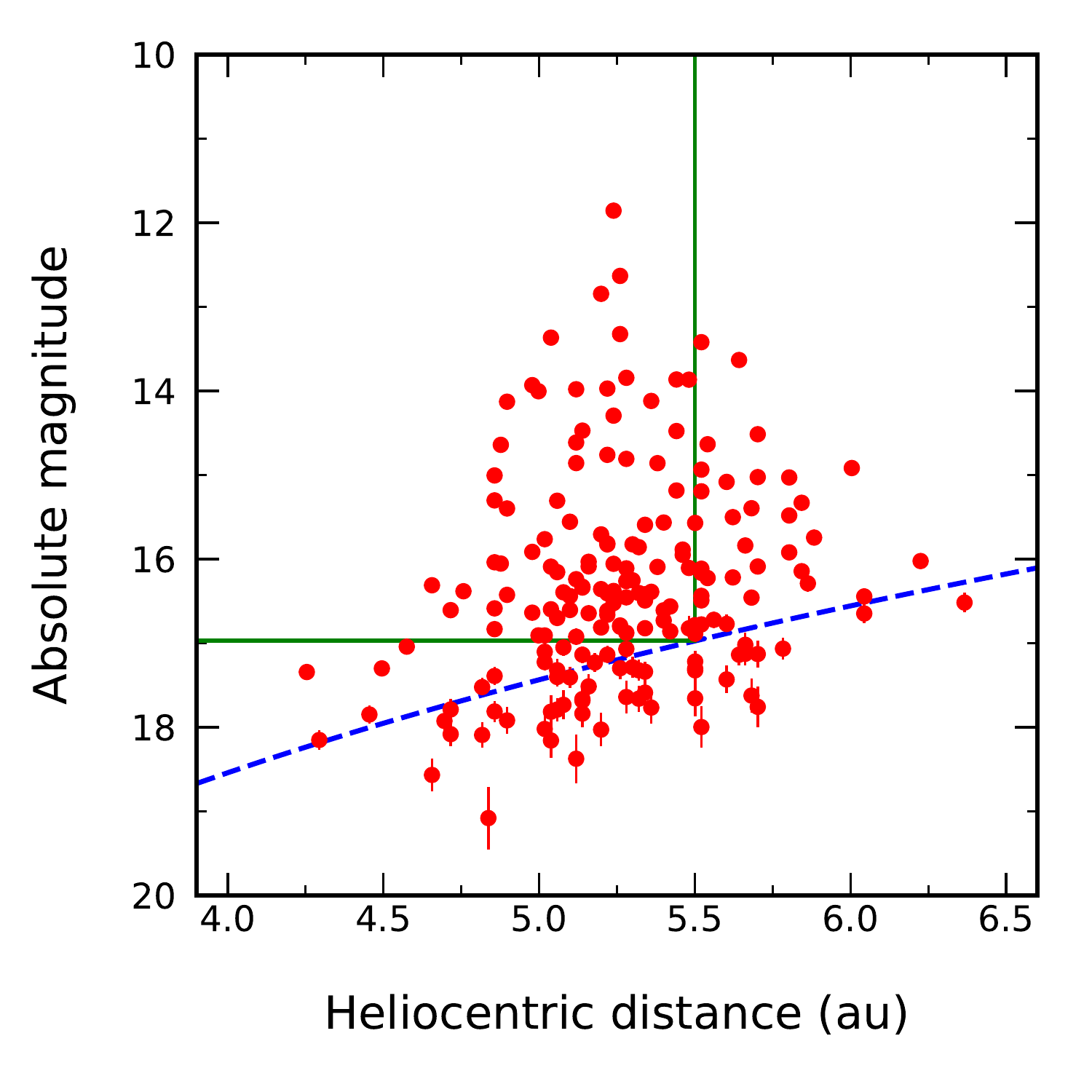} 
 \end{center}
 \caption{
 Absolute magnitudes with error bars as a function of heliocentric distance for the 187 Trojan asteroids for which photometric measurements were successfully performed. 
 The dashed lines shows the limiting apparent magnitude of 24.1 mag.
 The solid lines represent the boundaries we set to select our unbiased sample, i.e., $R = 5.5$ au and $H_r = 17.0$.
} 
\label{fig:f5}
\end{figure}

Figure \ref{fig:f5} shows the plots of the absolute magnitudes as a function of the heliocentric distance for each of the 187 Trojans for which photometric measurements were performed.
The uncertainty in the absolute magnitude comes from errors in the estimates of the apparent magnitude as well as the heliocentric and geocentric distances.
In our observations, measurement errors in apparent magnitude were less than $\sim 0.05$ mag for objects with $m_r \leq 23$ mag, while they were as large as 0.15 mag for $m_r \simeq 24$ mag.
Also, errors in the estimated heliocentric and geocentric distances were about 0.09 au \citep{yt17}.
Taking all these into account, errors in the obtained absolute magnitude of each object are shown with error bars in Figure \ref{fig:f5}.

The dashed line in Figure \ref{fig:f5} represents the apparent magnitude of the detection limit,  $m_r = 24.1$ mag, derived in \S 2.
In order to remove detection bias caused by the decreasing observed brightness with increasing distance from the Sun and Earth, we defined the outer edge of the heliocentric distance of detected Trojans by $R=5.5$ au, where $m_r = 24.1$ mag corresponds to $H_r = 17.0$ mag, and selected objects that satisfy $R \leq 5.5$ au and $H_r \leq 17.0$ mag as our unbiased sample.
The extracted sample consists of 87 Jupiter Trojans.

\subsection{Size Distribution}\label{sec:distr}

The obtained absolute magnitude $H_r$ for each object can be converted into its diameter $D$ by \citep{pr07}

\begin{equation}
\log D = 0.2 m_{\odot,r} + \log (2R_\oplus) - 0.5 \log p - 0.2 H_r,
\end{equation}

\noindent
where 
$m_{\odot,r}$ is the apparent $r$-band magnitude of the Sun ($-26.91\;$mag; \citet{fu11}), $R_\oplus$ is the heliocentric distance of Earth (i.e. 1au) in the same units as $D$, and $p$ is the geometric albedo.
We assumed $p=0.05$ on the basis of a recent analysis of {\it Near-Earth Object Wide-field Infrared Survey Explorer} (NEOWISE) data for Jupiter Trojans with $D \geq 10$ km \citep{ro18}.

We derived absolute magnitude distribution from our unbiased sample consisting of 87 objects, which can be converted into their size distribution.
Here, the derived number of objects should be corrected with the detection efficiency.
Suppose that the detection efficiency at $i$-th visit ($i=1$, 2, 3 in our observations) for an object $j$ with its apparent magnitude $m_j$ is given by $\eta_i(m_j)$.
Then the efficiency for the detection of this object in all the three visits can be given as

\begin{equation}
\eta(m_j) = \prod_{i=1}^3 \eta_i (m_j).
\end{equation}

\noindent
Thus, the corrected cumulative number of objects with absolute magnitude smaller than $H_r$ is 

\begin{equation}
N(<H_r) = \sum_{j:H_j<H_r} \frac{1}{\eta(m_j)}.
\label{eq:ncorrected}
\end{equation}

\begin{figure}
\begin{center}
 \includegraphics[height=10cm]{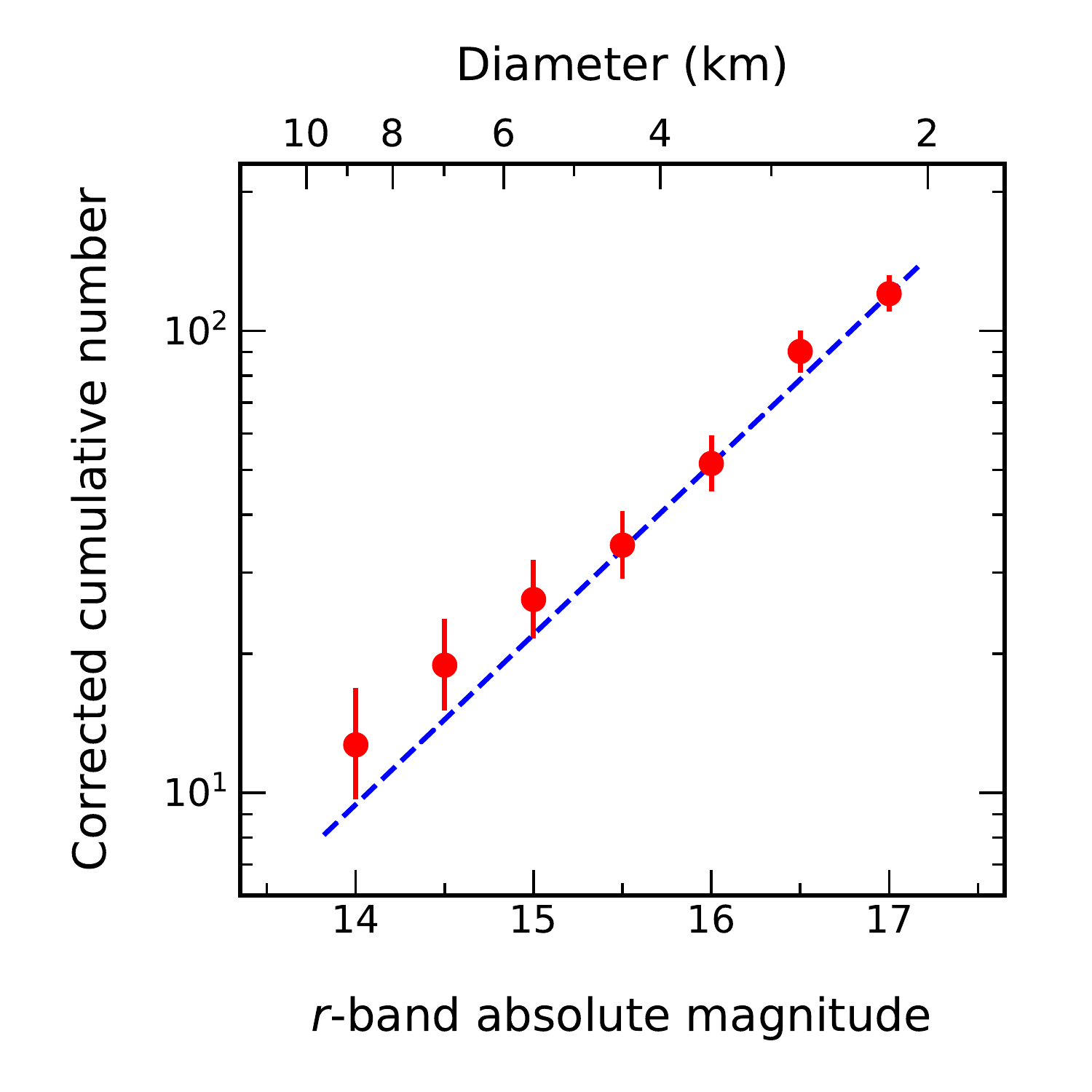} 
 \end{center}
 \caption{
 Cumulative absolute magnitude distribution for the 87 unbiased L${}_5$ Jupiter Trojans obtained from our survey (red dots with error bars).
 Note that the cumulative numbers are corrected using the detection efficiency as shown in Eq.(\ref{eq:ncorrected}). 
 Corresponding diameters for the assumed albedo of 0.05 \citep{ro18} are also shown at the upper horizontal axis.
 Best-fit single-slope power-law distribution is shown with the dashed line. 
} 
\label{fig:f6}
\end{figure}

\begin{figure}
\begin{center}
 \includegraphics[height=10cm]{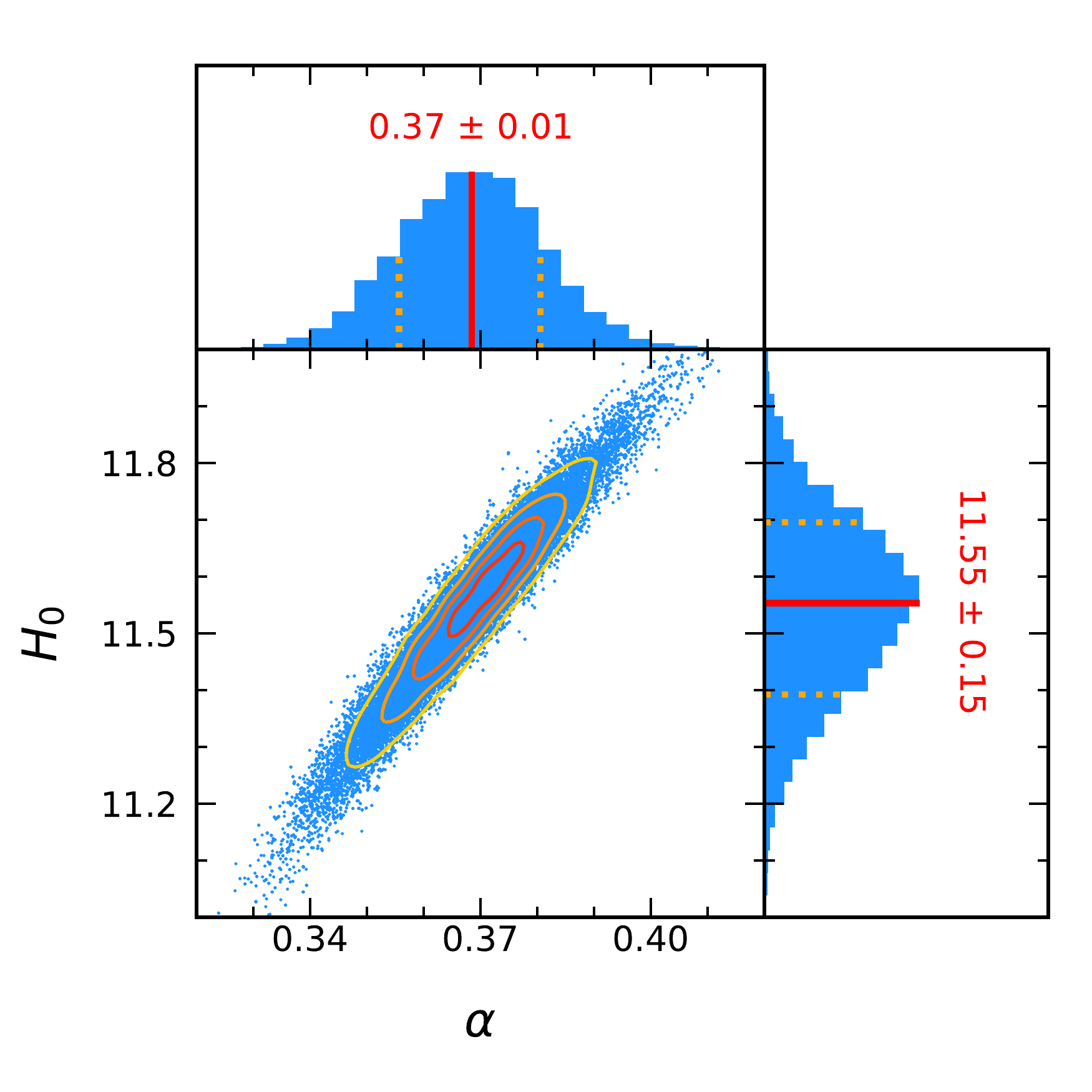} 
 \end{center}
 \caption{
Post-event distribution of sampling by MCMC for our unbiased L${}_5$ Jupiter Trojans.
 The red solid lines in the one-dimensional histograms show the best-fit values, and the orange dashed lines represent
 the resulting confidence intervals. Each blue point in the two-dimensional histograms and the contours show the results 
 of sampling for each step and their distributions, respectively. 
} 
\label{fig:f7}
\end{figure}

Figure \ref{fig:f6} shows the plots of the cumulative absolute magnitude distribution of the L${}_5$ Jupiter Trojans in our unbiased sample of 87 objects with $H_r < 17.0$ mag.
The corresponding diameter values are also indicated on the upper horizontal axis, which show that the Trojan asteroids in our sample have $D \simeq 2-10$ km.
We performed fitting of the distribution by a power-law with a single slope as follows.
The differential absolute magnitude distribution function defined by $\Sigma (H) = dN(< H_r)/dH$ for a single-slope power-law can be written as 

\begin{equation}
\Sigma (H) = 10^{\alpha(H-H_0)},
\label{eq:hpower}
\end{equation}

\noindent
where $\alpha$ is the power-law index and $H_0$ is defined so that $\Sigma (H_0) = 1$.
For a constant albedo, the cumulative diameter distribution can be written as

\begin{eqnarray}
N(>D) \propto D^{-b}
\end{eqnarray}

\noindent 
with $b=5\alpha$.

We used the maximum-likelihood method \citep[e.g.][]{be04} for fitting Eq.(\ref{eq:hpower}) to the obtained $H_r$ distribution.
For the numerical analysis of the maximum-likelihood estimation, we used the Markov Chain Monte Carlo method (MCMC) with a Python package \verb+emcee+ \footnote{https://emcee.readthedocs.io/en/stable/} \citep{fo13}.
Figure \ref{fig:f7} shows the post-event distributions of sampling for $\alpha$ and $H_0$ by MCMC.
We obtained the best-fit power-law index of $\alpha = 0.37 \pm 0.01$ (which corresponds to $b = 1.85 \pm 0.05$) with $H_0 = 11.55 \pm 0.15$, which is shown in Figure \ref{fig:f6} with the dashed line.
This obtained slope agrees with the result of \citet{yt17}, who also used data obtained using HSC and found $\alpha = 0.37 \pm 0.01$ for 431 small L${}_4$ Trojans in a similar size range.
This shows that the slope of the size distribution of L${}_4$ and L${}_5$ Trojans agree with each other for $D \simeq 2-10$ km.

\begin{figure}
\begin{center}
\includegraphics[height=10cm]{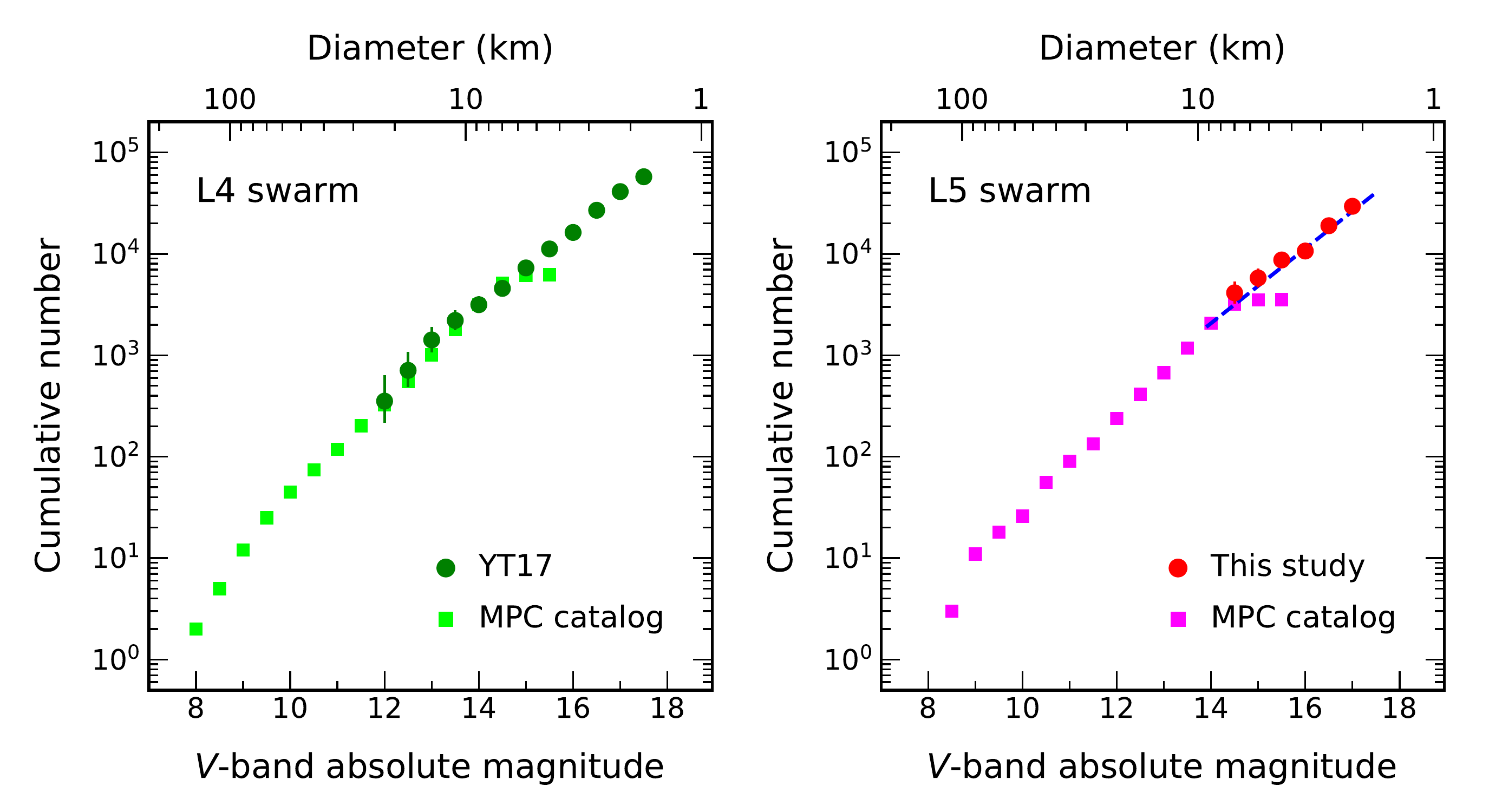}
\end{center}

\caption{
Left: Combined absolute magnitude distribution for the L${}_4$ swarm of the known Trojan asteroids taken from the MPC catalog (green squares) and those from the  HSC observation by \citet[][green circles]{yt17} scaled at $H_V = 14.0$ mag.
The $V-r$ color was assumed to be 0.25 mag \citep{sz07} in converting the HSC $r$-band magnitudes into the $V$-band magnitudes.
Corresponding diameters for the assumed albedo of 0.05 \citep{ro18}
are also shown at the upper horizontal axis. 
Note that \citet{yt17} assumed an albedo of 0.07 based on \citet{gr12}, while we used 0.05 for both swarms based on the above more recent work.
Right: Same as the left panel but for the L${}_5$ swarm.
The HSC data (red circles) in this case are those obtained by the present work, and are scaled so that their best-fit single-slope power-law distribution (dashed line; see also Figure \ref{fig:f6}) matches the MPC data (magenta squares) at $H_V = 14.0$ mag (see text).
}
\label{fig:f8}
\end{figure}


\begin{figure}
\begin{center}
\includegraphics[height=11cm]{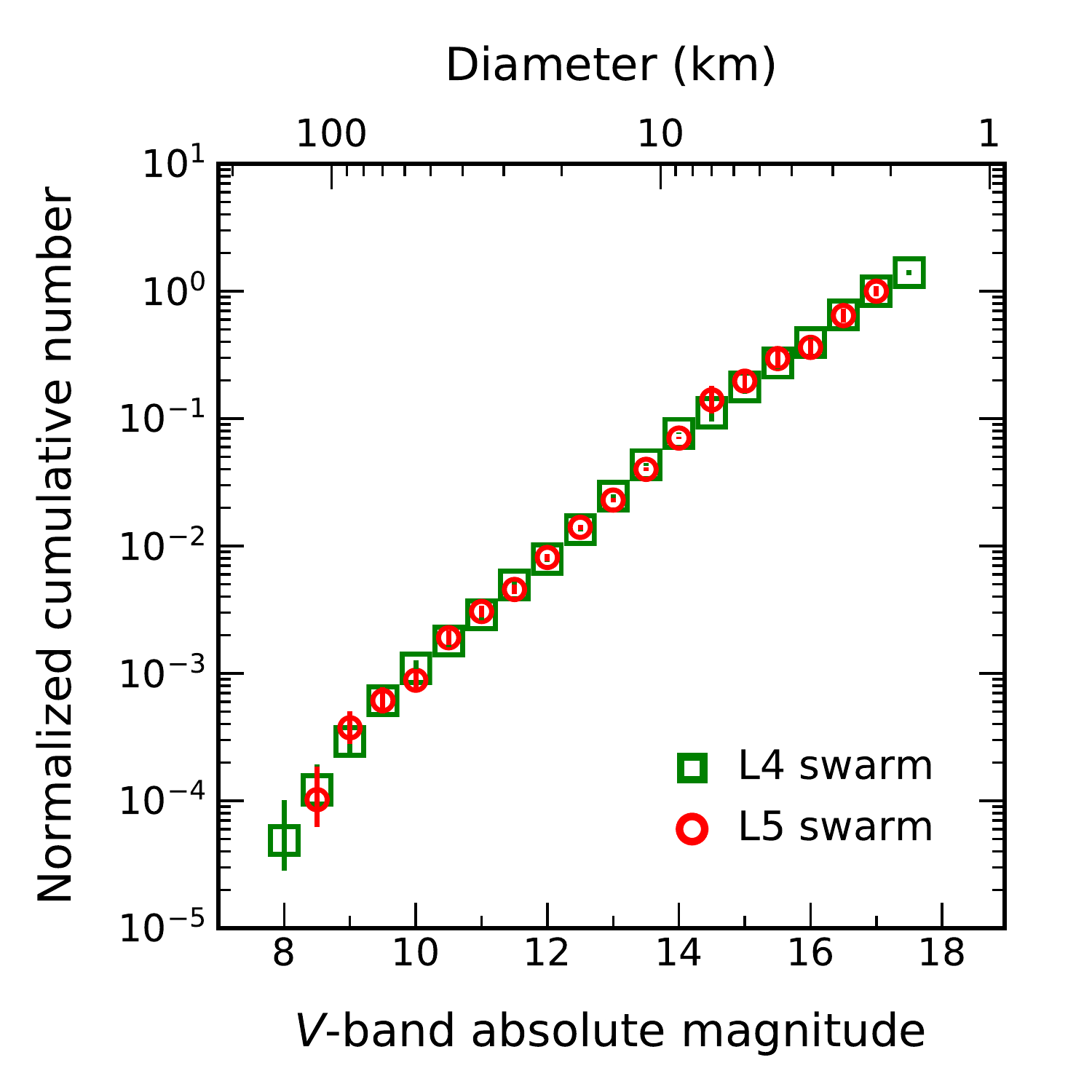}
\end{center}

\caption{
Combined absolute magnitude distribution for the L${}_4$ (squares) and L${}_5$ (circles) swarms normalized by their values at $H_V = 17.0$ mag.
}
\label{fig:f9}
\end{figure}


Next, we compare size distributions of L${}_4$ and L${}_5$ Trojans for a wider range of diameters, by combining the data obtained by HSC (\citet{yt17} and this work) with those for known Jupiter Trojans.
The left panel of Figure \ref{fig:f8} shows the distributions of $V$-band absolute magnitude and estimated diameters for L${}_4$ Trojans; the green circles show the data obtained by HSC \citep{yt17} and the green squares show the data for known Jupiter Trojans obtained from the MPC catalog.
In order to compare with the MPC data, which are given in $V$-band magnitude, we assumed the $V-r$ color to be 0.25 mag \citep{sz07} and scaled the distribution for the HSC data so that they match the MPC data at $H_V=14.0$ mag \citep{yt17}.
\citet{yt17} showed that the combined distribution can be fitted by a broken power-law with  $\alpha_1 = 0.50 \pm 0.01$ for $H_V \leq H_{\rm B}$  and $\alpha_2 = 0.37 \pm 0.01$ for $H_V > H_{\rm B}$ with break magnitude $H_{\rm B}  = 13.56^{+0.04}_{-0.06}$ mag.

The right panel of Figure \ref{fig:f8} shows similar plots for L${}_5$ Jupiter Trojans, where the HSC data obtained in the present work are combined with the data for known L${}_5$ Trojans taken from the MPC catalog.
We converted the $r$-band magnitude of our sample into the $V$-band magnitude in the same way as in \citet{yt17}.
Then we scaled the best-fit single-slope power-law distribution to our HSC data (Figure \ref{fig:f6}) so that it matches the MPC data at $H_V = 14.0$ mag, avoiding the use of the MPC data near small size end where the slope of the magnitude distribution begins to decay.
These plots show similarity of the size distributions of the two swarms for a wide range of sizes.
We calculated the slope of the magnitude distributions at $11.0 < H_V < 14.0$ using data given in the MPC catalog (updated April 23, 2021) by least-square fitting and found that 
$\alpha = 0.48 \pm 0.03$ for L${}_4$ and $\alpha = 0.46 \pm 0.03$ for L${}_5$, respectively.
Figure \ref{fig:f9} shows the combined cumulative magnitude distributions for the L${}_4$ and L${}_5$ swarms normalized at their values at $H_V = 17.0$ mag.
This clearly demonstrates that the size distributions of L${}_4$ and L${}_5$ Trojans agree well with each other in the size range of  $D \geq 2$ km.

By comparing the combined size distribution for the   L${}_5$ swarm based on our results (right panel of Figure \ref{fig:f8}) with that for the L${}_5$ swarm based on \citet[][left panel of Figure \ref{fig:f8}]{yt17}, we find that the ratio of the total numbers of L${}_4$ and L${}_5$ Trojan asteroids with $D \geq 2$ km is estimated to be $N_{\rm L4}/N_{\rm L5}=1.40 \pm 0.15$, which is consistent with previous works that obtained the ratio for larger asteroids \citep[e.g.][see \S \ref{sec:42}]{gr12}.
Extrapolating the newly obtained size distribution, the total number of L${}_5$ Jupiter Trojans with $D \geq 1$ km is estimated to be $1.1 \times10^5$, while L${}_4$ Jupiter Trojans with $D \geq 1$ km is estimated to be $1.5 \times10^5$ from the extrapolation of the results of \citet{yt17}.
Combining these, the total number of Jupiter Trojans (including both L${}_4$ and L${}_5$ swarms) with $D \geq 1$ km is estimated to be $2.6 \times 10^5$.
On the other hand, using data also obtained by the Subaru HSC, \citet{ma21} show that the total number of the main-belt asteroids with $D \geq 1$ km is about $2 \times 10^6$ (their Figure 10), although there is some uncertainty in the estimate.
This indicates that the total number of Jupiter Tronjans with $D \geq 1$ km is significantly (probably by nearly an order of magnitude) smaller than that of the main-belt asteroids in the same size range.

\section{DISCUSSION} \label{sec:disc}

\subsection{Comparison with Other Previous Studies}

We have shown that the results for our L${}_5$ Jupiter Trojan sample are consistent with those obtained by \citet{yt17} for the L${}_4$ swarm.
Here we compare our results with other previous works on the size distribution of Jupiter Trojans (Table \ref{table:t2}).
\citet{je00} derived size distribution of L${}_4$ Jupiter Trojans from observations using the University of Hawaii 2.2 m telescope, and obtained $\alpha = 0.40 \pm 0.05$ ($b = 2.0 \pm 0.25$) for $11 \lesssim H \lesssim 16$.
\citet{sz07} derived size distribution for the whole Jupiter Trojans including both L${}_4$ and L${}_5$ swarms using the SDSS Moving Object Catalog 3 (MOC3), and obtained  $\alpha = 0.44 \pm 0.05$ ($b = 2.2 \pm 0.25$) for $9 \lesssim H \lesssim 13.5$.
\citet{gr11} derived size distributions of L${}_4$ and L${}_5$ Jupiter Trojans using data obtained by the NEOWISE survey, 
and found that the distributions for the two swarms are similar with $\alpha = 0.4$ ($b = 2$) for $20 {\rm km} \lesssim D \lesssim 80$ km.
Using absolute magnitudes of Jupiter Trojans reported by the MPC, \citet{fr14} obtained  $\alpha_1 = 1.0 \pm 0.2$ and $\alpha_2 = 0.36 \pm 0.01$ for the large and small Jupiter Trojans, respectively, with a break at $H=8.4^{+0.2}_{-0.1}$.
\citet{wo14} examined size distributions of both L${}_4$ and  L${}_5$ Jupiter Trojans with $H < 12.3$ using the SDSS MOC4 data, and found that the two swarm distributions are indistinguishable within uncertainties.
Their results showed that $\alpha = 0.46 \pm 0.01$ for $H \geq 8.16$.
\citet{yt17} showed that their results are consistent with these previous works for the size range that overlaps with their HSC data.
Our results are also consistent with the above previous studies.
The similarity of the size distributions of the two swarms found by some of these previous works \citep[e.g.][]{gr11,wo14} is also consistent with the present work, which confirmed that similar size distributions are extended to smaller sizes.

Size distributions of small Jupiter Trojans ($D < 10$ km) have been investigated in detail using the Suprime-Cam CCD camera on Subaru Telescope.
Using a dataset obtained by a survey primarily targeting at main-belt asteroids \citep{yo03}, \citet{yn05} detected 51 L${}_4$ Jupiter Trojans and derived their size distribution.
They found that the observed size distribution can be fit either by a single slope power-law with $b=1.89$ ($\alpha = 0.378$) for $14 < H < 17.7$, or by a broken power-law with $b_1=2.39$ ($\alpha_1 = 0.48$) for $H < 16$ and $b_2=1.28$ ($\alpha_2 = 0.26$) for $H > 16$.
The power-law index for the above single-slope power-law distribution is consistent with the present work and \citet{yt17}.
Using a dataset obtained by another survey also primarily targeting at main-belt asteroids \citep{yn07}, \citet{yn08} detected 62 L${}_5$ Jupiter Trojans and derived their size distribution.
Their results showed that $b = 2.1 \pm 0.3$ ($\alpha = 0.42 \pm 0.06$) for $15.5 \lesssim H \lesssim 17.5$, but they did not find a break at $H \simeq 16$, which was found by \citet{yn05}.
Using HSC, which has a field of view about seven times as large as that of Suprime-Cam, \citet{yt17} and the present work performed surveys for the area of sky of about 26 deg${}^2$ and about 15 deg${}^2$, respectively; detected 631 and 189 Jupiter Trojans, respectively; and derived size distribution using unbiased samples consisting of 481 and 87 Trojans, respectively.
On the other hand, using Suprime-Cam, \citet{yn05} and \citet{yn08} carried out surveys for about 3 deg${}^2$ and about 4 deg${}^2$, respectively; detected 51 and 62 Jupiter Trojans, respectively; and derived size distribution using all these detected objects, but removal of observational bias may have not been sufficient.
Thus, it is possible that the difference in the size distributions found by these previous studies near the small size end such as the existence of a break at $H \simeq 16$ may be due to either the small number of detected objects and/or observational bias. 
More recently, \citet{wo15} carried out a survey of L${}_4$ Jupiter Trojans using Suprime-Cam, and obtained 
$b_0 \simeq 4.55$ ($\alpha_0 \simeq 0.91$) for $H < 8.46$, 
$b_1 \simeq 2.2$ ($\alpha_1 \simeq 0.44$) for $8.46 < H < 14.9$, and
$b_2 \simeq 1.8$ ($\alpha_2 \simeq 0.36$) for $H > 14.9$.
These results are roughly consistent with the present work and \citet{yt17}.
All these works show that the size distributions of small Jupiter Trojans are similar for the L${}_4$ and  L${}_5$ swarms, and can be approximated by
$b \simeq 4.5-5$ ($\alpha \simeq 0.9-1.0$) for $H \lesssim 8.5-9$;
$b \simeq 2.0-2.5$ ($\alpha \simeq 0.40-0.50$) for $9 \lesssim H < H_{\rm B}$; and 
$b \simeq 1.8-1.9$ ($\alpha \simeq 0.36-0.38$) for $H \gtrsim H_{\rm B}$,
with $H_{\rm B} \simeq 13.5-15$ (corresponding break diameter $\simeq 5-10$ km).


\begin{table}
\centering
\caption{Slope Indices and Break Magnitudes for Jupiter Trojan Absolute Magnitude Distribution}

\begin{tabular}{cccccc} \hline
\multirow{2}{*}{Reference} & \multirow{2}{*}{Swarm} & $\alpha_0$ & $\alpha_1$ & $\alpha_2$ & \multirow{2}{*}{$H_{\rm B0}$, $H_{\rm B}$} \\
                 &          & ($H < H_{\rm B0}$ ${}^{\rm (a)}$) & ($H_{\rm B0} < H < H_{\rm B}$ ${}^{\rm (b)}$) & ($H_{\rm B} < H$) &    \\ \hline
This work &   L${}_4$ &     & $0.48 \pm 0.03$ & & \\
               &   L${}_5$ &     & $0.46 \pm 0.03$ & $0.37 \pm 0.01$ & \\ 
\citet{yt17} &   L${}_4$ &     & $0.50 \pm 0.01$ & $0.37 \pm 0.01$ & $H_{\rm B}=13.56\ ^{+0.04}_{-0.06}$\\
\citet{wo15} &   L${}_4$ &  $0.91\ ^{+0.19}_{-0.16}$ ${}^{\rm (c)}$  & $0.45 \pm 0.05$ ${}^{\rm (c,d)}$& $0.36\ ^{+0.05}_{-0.09}$ ${}^{\rm (c)}$ & 
$\left\{ \begin{array}{c}
H_{\rm B0}=8.46\ ^{+0.49}_{-0.54} \\
H_{\rm B}=14.93\ ^{+0.73}_{-0.88}
\end{array} \right.$ \\ 
                                     &           &  $0.84\ ^{+0.22}_{-0.16}$  ${}^{\rm (e)}$& $0.33\ ^{+0.04}_{-0.03}$ ${}^{\rm (e)}$&  & 
$H_{\rm B0}=8.82\ ^{+0.35}_{-0.44}$ \\
                                      &        &  \multicolumn{2}{c}{$0.61\ ^{+0.07}_{-0.06}$  ${}^{\rm (f,g)}$} &  & 
$H_{\rm B0}=8.15\ ^{+0.06}_{-0.10}$\\
\citet{wo14} &   L${}_4$,  L${}_5$ &  $1.11 \pm 0.02$  ${}^{\rm (c)}$& $0.46 \pm 0.01$ ${}^{\rm (c)}$&  & 
$H_{\rm B0}=8.16\ ^{+0.03}_{-0.04}$ \\
                                     &           &  $0.97\ ^{+0.05}_{-0.04}$  ${}^{\rm (e)}$& $0.38 \pm 0.02$ ${}^{\rm (e)}$&  & 
$H_{\rm B0}=8.70\ ^{+0.08}_{-0.11}$ \\
                                      &        &  $1.25\ ^{+0.09}_{-0.04}$  ${}^{\rm (f)}$& $0.52\ ^{+0.03}_{-0.01}$ ${}^{\rm (f)}$&  & 
$H_{\rm B0}=8.15\ ^{+0.06}_{-0.10}$\\
\citet{fr14} &   L${}_4$,  L${}_5$ &  $1.0 \pm 0.2$ & $0.36 \pm 0.01$ &  & $H_{\rm B0}=8.4\ ^{+0.2}_{-0.1}$ \\
\citet{gr11} &   L${}_4$,  L${}_5$ &  & 0.4 &  & \\
\citet{yn08} &   L${}_5$ &  & & $0.42 \pm 0.06$ &  \\
\citet{sz07} &   L${}_4$,  L${}_5$ &  & $0.44 \pm 0.05$ &  & \\
\citet{yn05} &   L${}_4$ &  & & $0.38 \pm 0.02$ ${}^{\rm (h)}$&  \\
\citet{je00} &   L${}_4$,  L${}_5$ &  $0.89 \pm 0.15$ &  &  & $H_{\rm B0}=10.2 \pm 0.5$ \\
                 &   L${}_4$ &  & $0.40 \pm 0.05$ &  & \\
\hline
\end{tabular}
\tablecomments{
(a) $H_{\rm B0} \simeq 8 - 9$ \citep{fr14,wo14,wo15}, although an earlier work suggested a somewhat larger value \citep{je00}. 
(b) $H_{\rm B} \simeq 13.5 - 15$ \citep{wo15,yt17}.
(c) All sample. 
(d) This was obtained from fitting to their Subaru observation data, while  $\alpha_1 = 0.43 \pm 0.02$ was obtained from fitting to bright Trojans in the MPC catalog.
(e) Red objects only. (f) Less red objects only.
(g) The distribution for the less-red objects with $H < 12.3$ was found to be consistent with a single power law with this slope index.
(h) As an alternative distribution, the one with an additional break at $H \simeq 16$,
$\alpha = 0.48 \pm 0.02$ for $H \lesssim 16$ and  $0.26 \pm 0.02$ for $H \gtrsim 16$, was also suggested.
}
\label{table:t2}
\end{table}

On the other hand, it has been shown that the spectral slopes and albedos of Jupiter Trojans can be divided into two groups, a "red" group consistent with the asteroidal D-type and a "less-red" group consistent with the asteroidal P-type \citep[e.g.][]{sz07,em11,gr12,wo14}.
While the red-sloped spectra observed at visible and near-infrared wavelengths are often interpreted as indicating the existence of organic material, absence of strong absorptions expected for such material has been reported and the observed spectra may be explained by amorphous and/or space-weathered silicates \citep{eb03,eb04,em11,em15}.
\citet{wo14} examined the magnitude distributions of Jupiter Trojans in each of these two color populations with $H < 12.3$ and found that both the bright-end and the faint-end slopes are different (i.e., $\alpha^{\rm R}_1  = 0.97^{+0.05}_{-0.04}$ versus $\alpha^{\rm LR}_1  = 1.25^{+0.09}_{-0.04}$ for the bright-end and $\alpha^{\rm R}_2  = 0.38 \pm 0.02$ versus $\alpha^{\rm LR}_2  = 0.52^{+0.03}_{-0.01}$ for the faint-end, where "R" and "LR" denote red and less-red color populations, respectively).
\citet{wo15} performed observations of colors and magnitude distributions of 557 small L${}_4$ Jupiter Trojans using Suprime-Cam on Subaru Telescope with $g'$- and $i'$-filters.
In this case, most likely owing to the effect of asteroid rotation, the bimodality in color distribution that was seen for larger objects was not seen.
However, a tendency of decreasing $g$-$i$ color with increasing magnitude was confirmed, which is consistent with the difference in the slopes of magnitude distributions found by \citet{wo14}.
Such a difference in size distributions for the two color populations may reflect origins and collisional evolution of these bodies \citep{wo15,wo16}, and further observations as well as modeling efforts would be desirable.

\subsection{Implications for the Origin of Jupiter Trojans}
\label{sec:42}

The results of the present study as well as previous works on the size distribution of Jupiter Trojans show that the size distributions of the L${}_4$ and  L${}_5$ swarms agree well with each other for a wide range of sizes.
\citet{fr14} compiled Kuiper Belt survey data and compared the absolute magnitude distributions of hot and cold Kuiper-belt populations with Jupiter Trojans.
They found that, for the hot KBOs and Jupiter Trojans, the bright and faint object slopes and the break magnitude (when corrected for the bimodal albedo distribution for the hot KBOs) are statistically indistinguishable, while the cold classical KBOs show a much steeper large object slope.
\citet{fr14} argued that the similarity between the magnitude distributions of the hot KBOs and Jupiter Trojans is consistent with the scenario that they originated from the same primordial population and were scattered to their current locations afterwards \citep{mo05,ne13}.
The similarity between the size distributions of L${}_4$ and  L${}_5$ Jupiter Trojans that we confirmed in the present work is also compatible with the view that these Trojan asteroids originated from the same primordial population.

\begin{table}[h]
\centering
\caption{Degree of L${}_4/$L${}_5$ Asymmetry}

\begin{tabular}{cc} \hline
Reference & $N_{L4}/N_{L5}$ \\ \hline
This work &   $1.40 \pm 0.15$ \\
\citet{gr12} &  $1.34$ \\
\citet{gr11} &  $1.4 \pm 0.2$  \\
\citet{ny08} &   $1.9 \pm 0.4$ \\
\citet{sz07} &   $1.6 \pm 0.1$  \\
\hline
\end{tabular}

\label{table:t3}
\end{table}

On the other hand, it is known that the L${}_4$ and  L${}_5$ swarms have asymmetry in terms of total number (Table \ref{table:t3}).
\citet{sz07} estimated from the SDSS MOC 3 data that $N_{\rm L4}/N_{\rm L5}=1.6 \pm 0.1$.
On the basis of the number of detected Trojans for the two swarms by Subaru Suprime-Cam \citep{yn05,yn08} and a model for their surface number density distribution, \citet{ny08} estimated the total number of Trojans with $D > 2$ km in each swarm, which leads to $N_{\rm L4}/N_{\rm L5}=1.9 \pm 0.4$.
Using the data of Jupiter Trojans with $D > 10$ km seen by WISE, \citet{gr11} obtained $N_{\rm L4}/N_{\rm L5}=1.4 \pm 0.2$, while \citet{gr12} found $N_{\rm L4}/N_{\rm L5} \simeq 1.34$ also using the WISE data but for large objects with $D > 50$ km.
Our results show that $N_{\rm L4}/N_{\rm L5}=1.40 \pm 0.15$ for $D \geq 2$ km (\S \ref{sec:distr}), which is consistent with these previous works.
The model of capture of Trojans based on the original Nice model \citep{mo05} predicts equal efficiency of capture into the L${}_4$ and  L${}_5$ regions and cannot explain the observed asymmetry.
On the other hand, the model of capture based on the so-called "jumping Jupiter model" \citep{ne12} is potentially capable of producing an asymmetry by late passages of an ice giant near the L${}_5$ region that may have depleted the L${}_5$ population \citep{ne13}, and the model of capture due to the growth and inward migration of Jupiter can naturally explain the observed asymmetry \citep{pi19}.
Although the results of our present study cannot specify mechanisms that caused the asymmetry, our results suggest that the mechanism that lead to the asymmetry did not cause a noticeable difference between the size distributions of the two swarms.

Recently, \citet{he19} investigated the influence of the Yarkovsky effect on the long-term orbital evolution of Jupiter Trojans.
They found that objects with radii $R \lesssim 1$ km are significantly influenced by the Yarkovsky effect and these small bodies can be depleted over timescales of $\sim 10^8$ years, creating a shallower slope in the size distribution with a turning point at 100 m $\lesssim R \lesssim$ 1 km.
This is a size range below that of the currently observable population, thus we cannot test this model with the currently available observational data.
Future observations of sub-km Jupiter Trojans would provide further constraints on their origin and evolution.


\section{CONCLUSION} \label{sec:conclusion}

In the present work, using data obtained by HSC on Subaru Telescope, we derived size distribution of small Jupiter Trojans in the L${}_5$ swarm with $14 \lesssim H \lesssim 17$, corresponding to 2 km $\lesssim D \lesssim 10$ km.
We found that the obtained distribution can be approximated by a single-slope power-law, with an index for the differential magnitude distribution $\alpha = 0.37 \pm 0.01$, which corresponds to the cumulative size distribution for an assumed constant albedo with an index of $b = 1.85 \pm 0.05$.
This is consistent with \citet{yt17}, who examined the size distribution of small Jupiter Trojans in the L${}_4$ swarm also using HSC.
By combining our results with the size distribution of known L${}_5$ Trojans, we also obtained the size distribution of L${}_5$ Jupiter Trojans over the entire size range for $9 \lesssim H_V \lesssim 17$.
We also found that the slopes of the size distributions of L${}_4$ and  L${}_5$ Trojans agree well with each other at $H \gtrsim 9$, and their power-law indices are approximately given by $b \simeq 2.0-2.2$ ($\alpha \simeq 0.40-0.44$) for $9 < H < H_{\rm B}$ and $b \simeq 1.8-1.9$ ($\alpha \simeq 0.36-0.38$) for $H \gtrsim H_{\rm B}$
with $H_{\rm B} \simeq 13.5-15$ (corresponding break diameter $\simeq 5-10$km).
Extrapolating the newly obtained size distribution, the total number of L${}_5$ Jupiter Trojans with $D \geq 1$ km is estimated to be $1.1 \times 10^5$.
Combining this estimate with the result for the Jupiter Trojans in the L${}_4$ swarm in the same range \citep{yt17}, the total number of Jupiter Trojans with $D \geq 1$ km in the L${}_4$ and L${}_5$ swarms is estimated to be $2.6 \times 10^5$, which is significantly smaller than the recent estimate of the total number of the main-belt asteroids in the same size range \citep[$\sim 2 \times 10^6$;][]{ma21}.

It is known that the  L${}_4$ and  L${}_5$ swarms have asymmetry in their total number, with the L${}_4$ swarm having significantly more objects than the L${}_5$ swarm.
Combining our results for the L${}_5$ swarm with those obtained by \citet{yt17} for the L${}_4$ swarm, the ratio of the total number of asteroids with $D \geq 2$ km in the two swarms is estimated to be $N_{\rm L4}/N_{\rm L5}=1.40 \pm 0.15$.
Our results suggest that the mechanism that caused this asymmetry did not cause a noticeable difference between the size distributions of the two swarms.
Future observations of still smaller asteroids with larger telescopes and/or detailed close observations of asteroid surfaces by the Lucy spacecraft are expected to provide further constraints on the origin and evolution of Jupiter Trojans and the Solar System.


\acknowledgments
We thank Fumihiko Usui for helpful comments on an earlier draft of the manuscript, and the reviewers for comments and suggestions that helped improve the clarity of the manuscript. 
This work was supported by JSPS KAKENHI grant Nos. JP15H03716, JP16H04041, JP18K13607, JP20H04617, and JP21H00043.

\end{document}